\documentstyle[epsf]{elsart}
\def\dblspc {\smallskipamount=7.5pt plus2pt minus2pt
                  \medskipamount=15pt plus4pt minus4pt
                  \bigskipamount=30pt plus8pt minus8pt
                  \normalbaselineskip=30pt plus0pt minus0pt
                  \normallineskip=2pt
                  \normallineskiplimit=0pt
                  \jot=7.5pt
                  {\def\smallskip {\vskip\smallskipamount}}
                  {\def\medskip   {\vskip\medskipamount}}
                  {\def\bigskip   {\vskip\bigskipamount}}
                  {\setbox\strutbox=\hbox{\vrule
                    height21.0pt depth9.0pt width 0pt}}
                  \parskip 15.0pt
                  \normalbaselines}
\begin{document}
\dblspc
\begin{frontmatter}
\title{\bf Cumulative Dragging - An Intrinsic Characteristic of Stationary
Axisymmetric Spacetime}
\author{A. R. Prasanna\thanksref{email1}} and
\author{Sai Iyer\thanksref{email2}}
\thanks[email1]{E-mail: prasanna@prl.ernet.in}
\thanks[email2]{E-mail: sai@prl.ernet.in}
\address{Physical Research Laboratory, Ahmedabad 380 009, India}
\date{\today}

\begin{abstract}
\dblspc
The {\it Cumulative Drag Index\/} defined recently by Prasanna 
\cite{prasanna} has been generalised to include the centrifugal 
acceleration. We have studied the behaviour of the drag index for 
the Kerr metric and the Neugebauer-Meinel metric representing
a self-gravitating rotating disk and their Newtonian approximations.
The similarity of the behaviour of the index for a given set of
parameters both in the full and approximated forms, suggests that
the index characterises an intrinsic property of spacetime with
rotation. Analysing the index for a given set of parameters shows
possible constraints on them.
\end{abstract}
\end{frontmatter}

\section{Introduction}

The phenomenon of rotation plays a very important role in almost
all classes of objects that encompass our physical universe.
Particularly in the discussion of Inertia, rotational features
characterise global effects on local physics as implied by 
Mach's principle.  
Recently,  Prasanna \cite{prasanna} has defined a new parameter called
the {\em Cumulative Drag Index\/} for stationary  axisymmetric
spacetimes, using the 
notion of inertial forces within the framework of general
relativity.  The index, defined for particles in circular orbit
along the trajectory on which the centrifugal acceleration is
zero, characterises the intrinsic feature of rotation through the
drag induced on both co-rotating and counter-rotating particles. However, for
practical applications, it would be useful to generalise the drag
index to include the centrifugal acceleration also.  A few years
ago, astronomers discovered two co-spatial stellar disks in
the galaxy NGC-4550, one orbiting prograde and the other
retrograde with respect to the galactic nucleus, in the core of
the Virgo cluster \cite{rubin,rix}.
Bicak and Ledvinka \cite{bicak} have tried to 
construct sources for the Kerr geometry using counter-rotating
thin disks.  If one considers the galactic nucleus as a black
hole, one can then use the Kerr geometry for the outside and have
counter-streaming jets outside the ergo-region.  The presence of
such co- and counter-rotating particle streams may perhaps be
characterised through the drag index, defined as
\begin{equation}
\label{eq:cdigen}
{\cal C} = \frac{({\cal F}_{\rm cf}
                 +{\cal F}_{\rm co} 
                 -{\cal F}_{\rm gr})}
                {({\cal F}_{\rm cf}
                 +{\cal F}_{\rm co} 
                 +{\cal F}_{\rm gr})} \,,
\end{equation}
where ${\cal F}_{\rm cf}$, ${\cal F}_{\rm co}$, and ${\cal F}_{\rm
  gr}$, denote, respectively, the centrifugal, the Coriolis and the
gravitational accelerations acting on a particle in circular orbit in
a stationary, axisymmetric gravitational field.  Within the framework
of general relativity this definition is unique, when one considers
the spacetime expressed in the conformal 3+1 splitting with the four
acceleration $a_i$ being expressible covariantly as
\cite{abramowicz,prasanna}
\begin{equation}
\label{eq:accl}
a_i = -\nabla_i\phi + \gamma^2 V ( n^k \nabla_k \tau_i + \tau^k
\nabla_k n_i ) + (\gamma V)^2 \tilde{\tau}^k \tilde{\nabla}_k \tilde{\tau}_i\,.
\end{equation}
The various quantities on the r.h.s.\ of eq.~(\ref{eq:accl}) are as
described below: $n^i$ is the vector field corresponding to the
zero angular momentum observers expressed in terms of the
Killing vectors $\eta^i$ (timelike) and $\xi^i$ (spacelike) as
\begin{equation}
n^i = e^\phi( \eta^i + \omega\xi^i)\,,\qquad 
\omega = -\langle\eta,\xi\rangle / \langle \xi , \xi \rangle \,,
\end{equation}
and $\phi$ is the scalar potential
\begin{equation}
\phi = - \frac{1}{2}\ln (-\langle\eta,\eta\rangle - 2\omega\langle\xi,\eta\rangle 
-\omega^2 \langle\xi,\xi\rangle )\,.
\end{equation}
$\tau^i$ is the unit spacelike vector orthogonal to $n^i$ along
the circle depicting the orbit of the particle with a constant
speed $V$, and $\gamma$ ($ = 1/\sqrt{1-V^2}$) is the Lorentz factor.
The particle four velocity, $U^i$, is
thus expressible as 
\[
U^i = \gamma (n^i + V \tau^i ) \,,
\]
and is also equal to $A ( \eta^i + \Omega \xi^i )$, with $A$
the redshift factor defined as
\begin{equation}
A^2 = - ( \langle\eta,\eta \rangle + 2 \Omega \langle\xi,\eta\rangle + 
\Omega^2 \langle\xi,\xi\rangle )^{-1}\,,
\end{equation}
$\Omega$ being the angular velocity, 
$ \Omega\tau^i = e^\phi(\Omega-\omega) \xi^i$.  
$\tilde{\tau}^i = e^{-\phi} \tau^i$ is the
vector defined on the conformally projected 3-space having the
positive definite metric $h_{ik} = g_{ik} + n_i n_k$ and
$\tilde{\nabla}_i$ is the covariant derivative with respect to
$\tilde{h}_{ik} = e^{2\phi}h_{ik}$.

As shown earlier, for the metric
\begin{equation}
ds^2 = ( g_{tt} dt^2 + 2g_{t\phi} dt d\phi + g_{\phi\phi}d\phi^2 )
 + ( g_{rr} dr^2 + g_{\theta\theta}d\theta^2 )
\end{equation}
the accelerations are given as
\begin{eqnarray}
\mbox{\em Gravitational:\hspace{0.5in}} ({\cal F}_{\rm gr})_i 
&=& -\nabla_i\phi = \frac{1}{2}\partial_i \left\{ 
    \ln\left[\frac{g^2_{t\phi}-g_{tt}g_{\phi\phi}}{g_{\phi\phi}}\right]
\right\}\,, \label{eq:fgr}\\ [2em]
\mbox{\em Coriolis:\hspace{0.5in}}({\cal F}_{\rm co})_i 
&=& \gamma^2 V n^j (\nabla_j\tau_i - \nabla_i\tau_j)\nonumber\\
&=& -A^2 (\Omega-\omega) g_{\phi\phi} \partial_i(g_{t\phi}/g_{\phi\phi})
    \,,\label{eq:fco}\\ [2em]
\mbox{\em Centrifugal:\hspace{0.5in}}({\cal F}_{\rm cf})_i 
&=& (\gamma V)^2 \tilde{\tau}^k\tilde{\nabla}_k\tilde{\tau}_i\nonumber\\
&=& -\frac{A^2(\Omega-\omega)^2}{2} g_{\phi\phi} \partial_i \left\{ 
\ln\left[\frac{g^2_{\phi\phi}}{g^2_{t\phi}-g_{tt}g_{\phi\phi}}\right]
\right\}\,\label{eq:fcf}.
\end{eqnarray}

\section{Kerr spacetime}

Taking now the specific example of Kerr spacetime
\begin{eqnarray}
ds^2 =& -\left(1-\frac{2mr}{\Sigma}\right)dt^2
        - \frac{4mra}{\Sigma}\sin^2\theta dt d\phi
        + &\frac{B}{\Sigma}\sin^2 d\phi^2 \nonumber\\
&&\qquad +\frac{\Sigma}{\Delta}dr^2 + \Sigma d\theta^2,
\end{eqnarray}
where $B=(r^2+a^2)^2-\Delta a^2\sin^2\theta$,
$\Delta=r^2+a^2-2mr$ and $\Sigma=r^2+a^2\cos^2\theta$,
and considering a particle in circular orbit on the equatorial
plane $(\theta=\pi/2)$, it can be seen that the index is
\begin{eqnarray}
\label{eq:cdikerr}
{\cal C}&=&\biggl\{m[r^4(r-2m)+2a^2r(r^2-8mr+10m^2)+a^4(r-6m)]\nonumber\\
&&+2am\Omega[r^4(r+2m)+2a^2r(r^2+4mr-10m^2)+a^4(r+6m)]\nonumber\\
&&+\Omega^2(r^3+a^2r+2ma^2)[r^4(r-4m)+a^2r(r^2-5mr+10m^2)-3a^4m]
\biggr\}\nonumber\\
&&\biggl/\biggl\{\Delta(r^3+a^2r+2a^2m)(-m(1-a\Omega)^2+\Omega^2r^3)\biggr\}\,.
\end{eqnarray}
It is clear that, of the two infinities of the index, one appears
at the event horizon $(\Delta = 0)$ while the other depends on
both $a$ and $\Omega$ and appears for a given $a$ and $\Omega$
at $r = [m(1-a\Omega)^2/\Omega^2]^{1/3}$.  Fig.~(\ref{fig:cdi-om}) shows
the nature of $\cal C$ at the three locations, $r_{\rm php}$ (the
prograde photon orbit), $r_{\rm phr}$ (the retrograde photon
orbit) and $r_{\rm cfo}$ (the orbit where the centrifugal force
is zero).

While at $r_{\rm php}$ the index is positive only for a very
small range of $\Omega$ for counter-rotating particles $(\Omega
< 0)$, at $r_{\rm phr}$ the index is positive for the same
range of $\Omega$ for co-rotating particles only.  On the other
hand, as was discussed earlier in \cite{prasanna}, at $r_{\rm cfo}$
the index is positive for both co- and counter-rotating
particles, but again for a very narrow range of values of
$\Omega$ (Fig.~(\ref{fig:cdi-om}c)).  
This change of behaviour of ${\cal C}$ at
the two photon orbits arises due to the following reason:  When
the centrifugal force is not zero, the two zeros of the
denominator of $\cal C$ outside the event horizon corresponding to
the fixed value of $r$ and $a$, are at
\begin{equation}
\Omega_1 = \frac{\sqrt{m}}{a\sqrt{m} + r^{3/2}}\qquad\mbox{and}\qquad
\Omega_2 = \frac{\sqrt{m}}{a\sqrt{m} - r^{3/2}}\,,
\end{equation}
and corresponding to these two $\Omega$ values the numerator of
$\cal C$ factors as
\begin{equation}
2mr^{5/2} (-2 a \sqrt{m} + 3m r^{1/2} - r^{3/2}) [ 4a^2mr 
- ( a^2 + r^2 )^2]
\end{equation}
and
\[
2mr^{5/2} ( 2a\sqrt{m} + 3mr^{1/2}  - r^{3/2} ) [
4a^2 mr - ( a^2 + r^2 )^2 ]\,,
\]
respectively.  Thus the zero at $\Omega_1$ cancels with the
numerator at the prograde photon orbit, while the one at
$\Omega_2$ cancels with the numerator at the retrograde photon
orbit. 

For many applications, often one takes the view that the
linearized Kerr metric might be sufficient to incorporate the
relativistic effects, when the body is slowly rotating.  In
order to examine this, let us consider the nature of $\cal C$
under this approximation. The three accelerations acting on a
particle in circular orbit, approximated to terms linear in the
Kerr parameter $a$, are given by
\begin{eqnarray}
\label{lforces}
{\cal F}_{\rm gr}&=& 
\frac{m}{r^2} \left(1 -\frac{2m}{r} \right)^{-1}\,,\nonumber\\ [2em]
{\cal F}_{\rm co}&=& 
-\frac{6am\Omega}{( r^2 - 2mr - \Omega^2
  r^4)}\,,\nonumber\\ [2em]
{\cal F}_{\rm cf}&=&
\frac{\Omega(r-3m)(\Omega^3r^6 - \Omega r^4 + 2 m \Omega r^3 
+ 4mar - 8am^2)}{r(r-2m)(r-2m-\Omega^2 r^3)^2}\,,
\end{eqnarray}
and thus the index is
\begin{eqnarray}
\label{lcdi}
\ell{\cal C}&=&\biggl[-m(r-2m)-2am\Omega r-\Omega^2r^3(r-5m)\nonumber\\
&&\qquad+6am\Omega^3r^3+\Omega^4r^6(r-4m)/(r-2m)\biggr]\biggl/\nonumber\\
&&\biggl[m(r-2m)-2am\Omega r-\Omega^2r^3(r-m)+6am\Omega^3r^3+
\Omega^4r^6\biggr]\,.
\end{eqnarray}
The very first change one notices is that, neglecting $a^2$ and
higher order terms in $a$, moves the infinity at the horizon to
$r = 2m$, as this would now represent the horizon, just like in
the static case.  Similarly, the orbit where the centrifugal
acceleration is zero also coincides with that in the static
case, viz., $r = 3m$, for all $\Omega$.  Fig.~(\ref{fig:lcdi-om}) shows the
index for the linearised version at the two photon orbits and at
the orbit on which the centrifugal acceleration is zero.
Comparison of these plots with those for the unapproximated
$\cal C$ (Fig.~(1)) clearly shows that for particles with angular
velocity $|\Omega| > 0.3$, the behaviour is exactly same
with or without the approximation at the photon orbits whereas
at the orbit with ${\cal F}_{\rm cf}=0$, the similarity is striking for 
all values of $\Omega$.  

As it would be almost impossible for particles to have a low
value of $\Omega$ close to photon orbits (as they would be
relativistic), the behaviour of the index shows that the
linearisation approximation for the forces is amply justified
for all practical purposes.  However, as a matter of principle
one finds that for very low values of $\Omega$ the behaviour of
the linearised version differs from that for the exact version,
the difference arising mainly because of the centrifugal
acceleration being non-zero.  If one considers the
behaviour of $\ell\cal C$ as a function of $r$ for fixed $a$ and
$\Omega$, it seems to be exactly like for $\cal C$, the expression
without approximation.

Looking at the overall feature of the inertial accelerations it
then seems that for understanding the frame dragging coming from
rotation, for practical purposes of considering the forces, it
may indeed be sufficient to calculate the gravitational,
Coriolis and centrifugal acceleration in the linearised
approximation, as given in eq.~(\ref{lforces}).

It is further interesting to consider the Newtonian limit of
the accelerations with lowest order corrections in the
centrifugal acceleration, as given by:
\begin{eqnarray}
{\cal F}_{\rm gr} &=& \frac{m}{r^2},\qquad 
{\cal F}_{\rm co} = - \frac{6am\Omega}{r^2}\,\nonumber \\
{\cal F}_{\rm cf} &=& - \Omega^2 (r-3m) + \frac{4am\Omega}{r^2} \,.
\end{eqnarray}
With these expressions the drag index turns out to be
\begin{equation}
N{\cal C} = \frac{(r^3-3mr^2)\Omega^2 + 2am\Omega + m}
{(r^3-3mr^2)\Omega^2 + 2am\Omega - m }\,.
\end{equation}
Fig.~(\ref{fig:ncdi-om}) shows the plots of $N\cal C$ as a 
function of $\Omega$, for
fixed $r$ and $a$ (\ref{fig:ncdi-om}a,b) and as a function of $r$ for
fixed $a$ and $\Omega$ (\ref{fig:ncdi-om}c,d).  
As the Newtonian approximation can be
valid only for larger values of $r$, it is clear that the index
is positive for both co- and counter-rotating particles for
$|\Omega|>0.1$, independent of the values
of $a$.

\section{Disk spacetime}

We shall next consider the behaviour of the index for the
Neugebauer-Meinel metric \cite{neug} for a rigidly rotating disk of
dust as given in the discussion of its dragging effects by
Meinel and Kleinw\"achter \cite{meinel}.  The metric components and
their first radial derivatives of significance are
\begin{eqnarray}
  \Omega_d^2 g_{\phi\phi}&=& \frac{\mu}{2} 
    -\left( \frac{z}{1+z} \right)^2\,, \quad
  \Omega_d g_{\phi t} = \frac{z}{1+z} - \frac{\mu}{2}\,,\quad
    g_{tt}= \frac{\mu}{2} -1\,,\nonumber\\
  \Omega_d^2 r_0 g_{\phi\phi,r}&=& \mu \frac{1-z}{1+z}\,,\quad
    \Omega_d r_0g_{\phi t,r}= \mu \frac{z}{1+z}\,,\quad
    r_0g_{tt,r}= -\mu \,\label{eq:diskmetric}.
\end{eqnarray}
where $\Omega_d$ is the angular velocity of the disk, 
$z$ represents the relative redshift of photons from the
centre of the disk measured at infinity, $\mu$ is a parameter
defined through the relation $\mu = 2\Omega_d^2 r^2_0 (1+z)^2$ and
$r_0$ represents the rim of the disk.
Using eqs.~\ref{eq:cdigen}, \ref{eq:fgr}--\ref{eq:fcf} and \ref{eq:diskmetric},
the drag index may be obtained as:
\begin{eqnarray}
\label{eq:dcdi}
{\cal C} &=& \biggl[
  \mu^2(\Omega_d^2-\Omega^2)+2\mu\{4(\mu-1)\Omega_d^2-5\mu\Omega_d\Omega
  + \mu\Omega^2\}z\nonumber\\
&&  \quad +2\{(4-15\mu+9\mu^2)\Omega_d^2 + 3\mu(4-5\mu)\Omega_d\Omega
  + \mu(6\mu-1)\Omega^2\}z^2\nonumber\\
&&  \quad 2\{2(4-9\mu+4\mu^2)\Omega_d^2-(8-26\mu+15\mu^2)\Omega_d\Omega
  + \mu(7\mu-8)\Omega^2\}z^3\nonumber\\
&&  \quad +(\Omega-\Omega_d)^2(8-14\mu+5\mu^2)z^4\biggr]\biggl/\nonumber\\
&&\biggl[ \mu(\Omega_d-\Omega) (1+z) \{\Omega_d(1+z)
+\Omega(1-z)\}\{\mu(1+z)^2-2z^2\}\biggr]\,,
\end{eqnarray}
where $\Omega$ is the angular velocity of the particle.

It is clear from the above expression that the two zeros of the
denominator, where the index blows up, correspond to the two
{\em circular geodesic orbits}, as the sum of the forces acting on
the particle is zero for these parameters.  While the zero at
$\Omega = \Omega_d$, corresponds to the prograde geodesic, the
one at $\Omega = -\Omega_d (1+z)/(1-z)$, corresponds to
the retrograde geodesic as also shown by Meinel and
Kleinw\"achter.  However, one finds that if $z > 1$, the second
zero occurs at a positive value of $\Omega$ as shown in the
Fig.~(\ref{fig:dcdi-om}d).  Fig.~(\ref{fig:dcdi-om})  shows the plots of 
the index for four different values of $\mu$.

\section{Discussion}

The presence of centrifugal acceleration does bring in a difference in
the behaviour of the drag index at the two photon orbits, with the
co-rotating ones having a positive value for a narrow range of
$\Omega$ at the retrograde photon orbit and the counter-rotating ones
having a similar feature at the prograde photon orbit.  However, if the
black hole is slowly rotating ($a\ll 1$, $a^2$ negligible), then
adopting the linearised version of the acceleration changes the
behaviour of the index only for very low values of $\Omega$
($|\Omega|< 0.3$) at the photon orbits, whereas for higher values of
$\Omega$ the behaviour resembles that of the full $\cal C$ without any
approximation.  On the other hand, for given $a$ and $\Omega$, as a
function of $r$ the radial distance parameter, the index shows no
change with approximation, thus indicating that the drag index
signifies something intrinsic to the spacetime with rotation, as its
behaviour for both co- and counter-rotating particles appears similar,
from the point of view of a locally non-rotating observer.

In order to appreciate the significance of the index defined, one can
consider its behaviour for the physical case of a rotating disk as
depicted in Fig.~(\ref{fig:dcdi-om}).  Comparing with the discussions of
Meiner and Kleinw\"achter, we find that the constraints on the value
of $z$ appearing through the parameter $\mu$, is well reflected in
Figs.~(\ref{fig:dcdi-om}a,b,c).  While Fig.~(\ref{fig:dcdi-om}a) shows
clearly the existence of two geodesic orbits (pro and retro) for
$\mu=0.1$, Fig.~(\ref{fig:dcdi-om}b) shows for $\mu = 0.5$, the existence
of only a prograde geodesic.  In fact it is instructive to compare
this with Fig.~(\ref{fig:cdi-om}b) which corresponds to ${\cal C}$ at the
retrograde photon orbit for Kerr, and thus conclude that $\mu = 0.5$
corresponds indeed to the last possible retrograde geodesic orbit.
Further, from the zeros of the denominator of ${\cal C}$
(eq.~\ref{eq:dcdi}), the appearance of the retrograde geodesic
corresponds to the particle angular velocity $\Omega = -\Omega_d
(1+z)/(1-z)$.  In fact, for this value of $\Omega$ the numerator
of the index factors as
\begin{equation}
8(2\mu-1)\Omega_d^2z(1+z)^2(\mu-z-\mu z)/(1-z)^2\,,
\end{equation}
which clearly shows that the zero of the denominator cancels with the term
$(2\mu-1)$ in the numerator for $\mu=1/2$, exactly similar to what happens
in the case of the Kerr metric. It is obvious that for the case $z>1$, 
i.e., $\mu>1.3519$, this value of $\Omega$ becomes positive, the infinity
of the index appearing for a prograde orbit.
However, this orbit turns out to be spacelike
($v>c$). Thus, it appears that the counter-streaming particles would
have a limitation from the point of view of their redshift and this
could play an important role in the analysis of particle motion in
the disk associated with the Virgo cluster.
From this discussion it
follows that given any stationary axisymmetric metric representing an
astrophysical situation, one can straightaway determine constraints on
possible physical parameters characterising the geodesic orbits
through the behaviour of the drag index.

Though Newtonian physics does not directly predict anything regarding
the nature of spacetime as influenced by rotation, it is amazing to
see that the cumulative drag index shows exactly similar behaviour in
the Newtonian approximation for the case of the full Kerr geometry
as well as for the Neugebauer-Meinel disk geometry
(Fig.~(\ref{fig:cdi-ncdi})), for all values of $a$ and $\Omega_d$,
either prograde or retrograde. 
Thus it is clear that the cumulative drag index defined
above characterises an intrinsic property of spacetime with rotation,
which goes beyond approximations.  The fact that it is positive for
both co- and counter-rotating particles having reasonable angular
velocities, outside the ergo-region of a black hole
or in a self-gravitating disk, clearly supports
the possibility of sustaining counter-rotating streams.
In fact, this analysis points out a constraint on the $z$ value for 
counterrotating streams to be $<0.285$, which might be tested in
the case of streams encountered in the Virgo cluster. Eventhough
the drag index itself does not measure any observable quantity
directly, it characterises intrinsic rotation for stationary,
axisymmetric spacetimes, be it in an empty region (like Kerr
geometry) or within a rotating disk (Neugebauer-Meinel class of
solutions), independent of approximations. If the metric
potentials depend upon a directly observable parameter then the
behaviour of the index for different combinations of rotational
parameters could possibly give constraints on the physical
parameter, which may be measurable.  Further, as the free orbits
are defined through the equilibration of the forces acting on a
particle, the index would go to infinity and thus studying the
behaviour of the index in general yields information for orbits
both free and otherwise.  Thus for the case of stationary
axisymmetric metrics, the location of geodesics for given
angular velocity and rotational parameters may be identified by
plotting the drag index without necessarily solving the
equations of motion. 

\section*{Acknowledgments}

We wish to thank Jayant Narlikar for raising the question of
the Newtonian limit of the drag index which led to the
investigations presented above, and the referees for some useful
comments.

\begin{figure}
\caption{$\cal C$ as a function of $\Omega$ ($a=0.5m$) at $r_{\rm php}$ (a), 
$r_{\rm phr}$ (b) and $r_{\rm cfo}$ (c).}
\label{fig:cdi-om}
\epsfxsize 3in
\centerline{\epsfbox{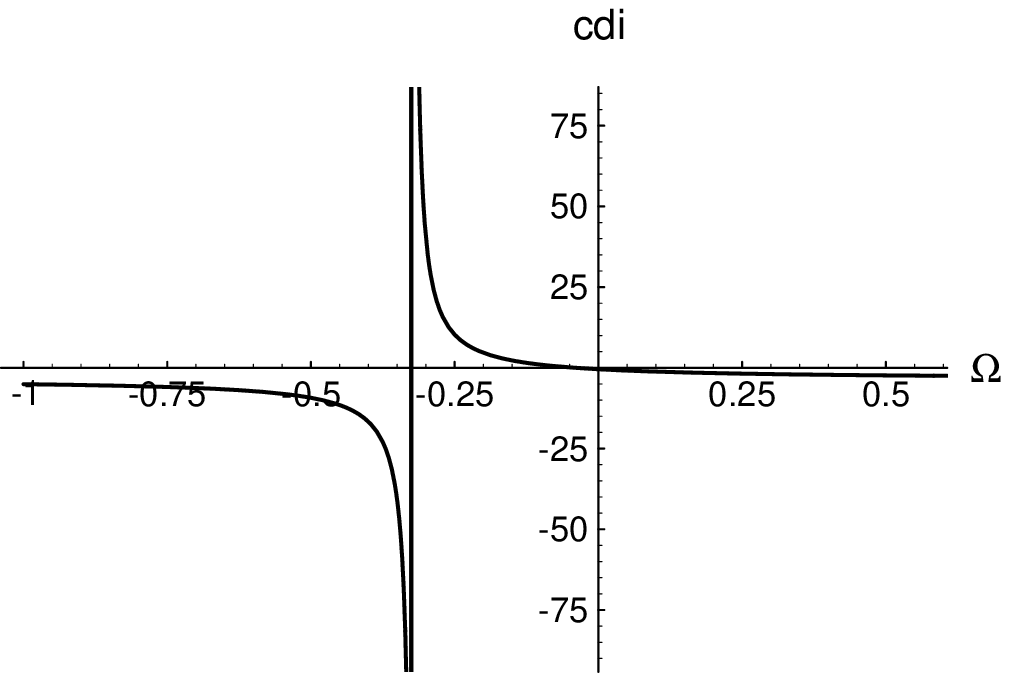}}
\centerline{(a)}
\epsfxsize 3in
\centerline{\epsfbox{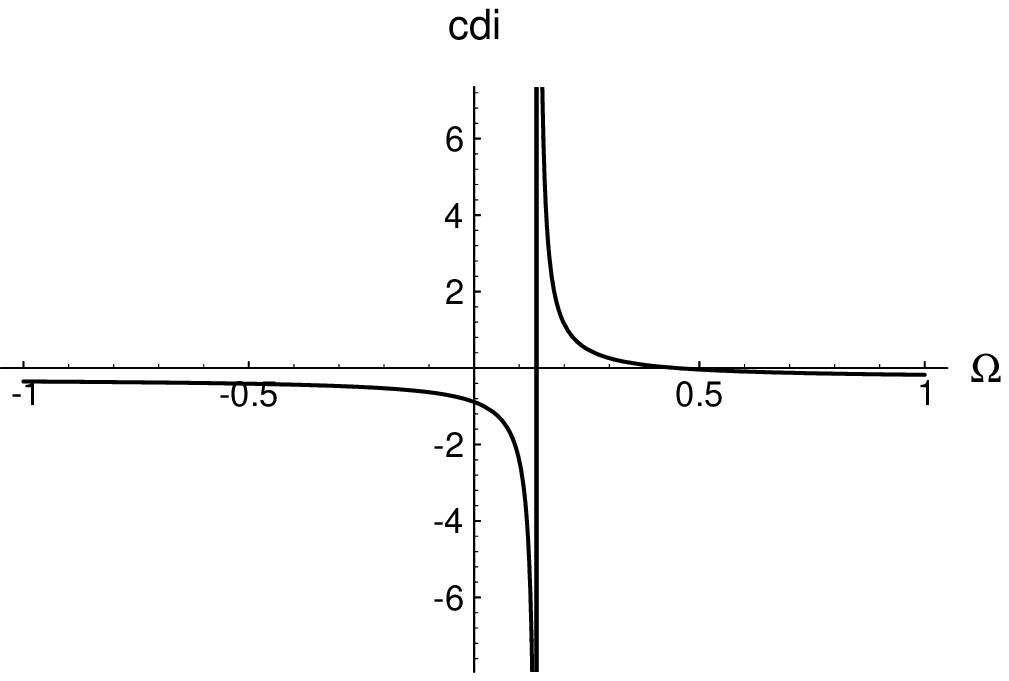}}
\centerline{(b)}
\epsfxsize 3in
\centerline{\epsfbox{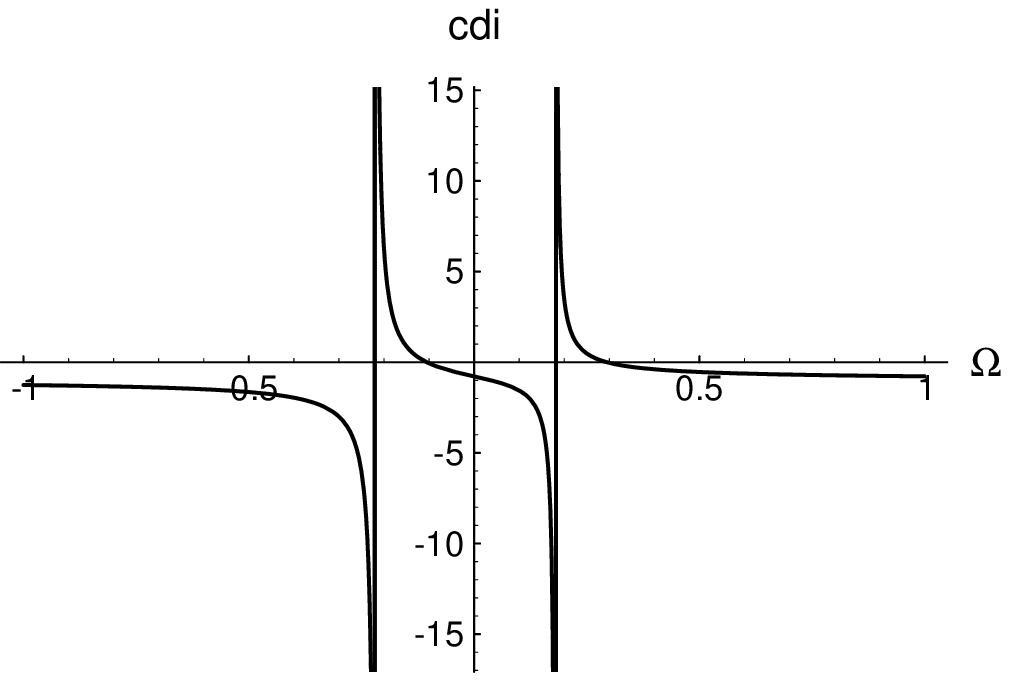}}
\centerline{(c)}
\end{figure}

\begin{figure}
\caption{$\ell\cal C$ as a function of $\Omega$ ($a=0.5m$) at 
$r_{\rm php}$ (a), 
$r_{\rm phr}$ (b) and $r_{\rm cfo}$ (c).}
\label{fig:lcdi-om}
\epsfxsize 3in
\centerline{\epsfbox{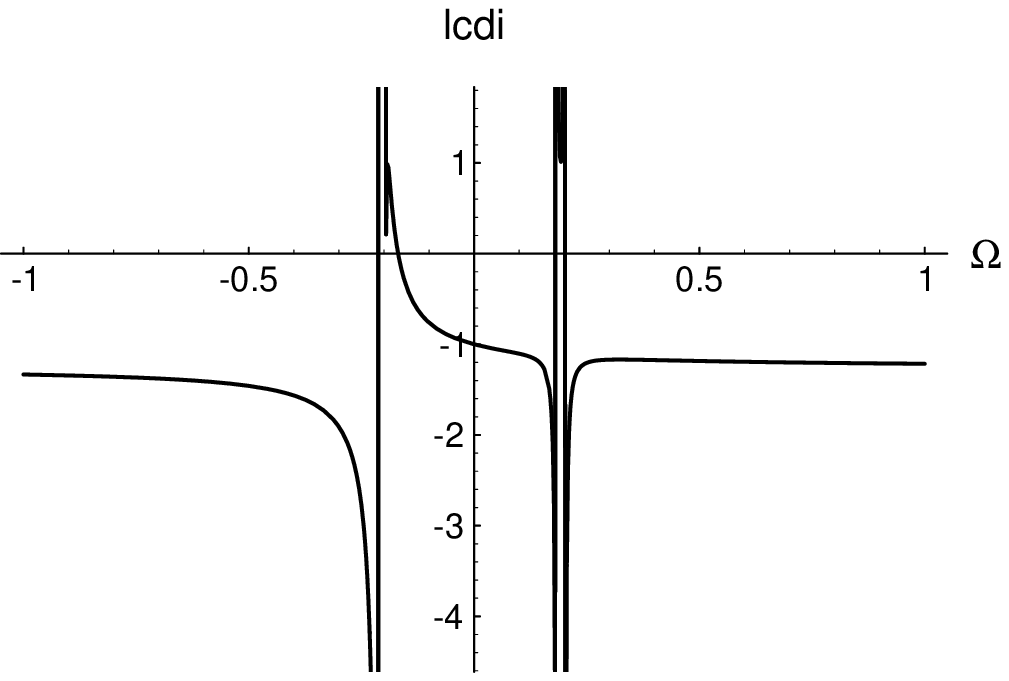}}
\centerline{(a)}
\epsfxsize 3in
\centerline{\epsfbox{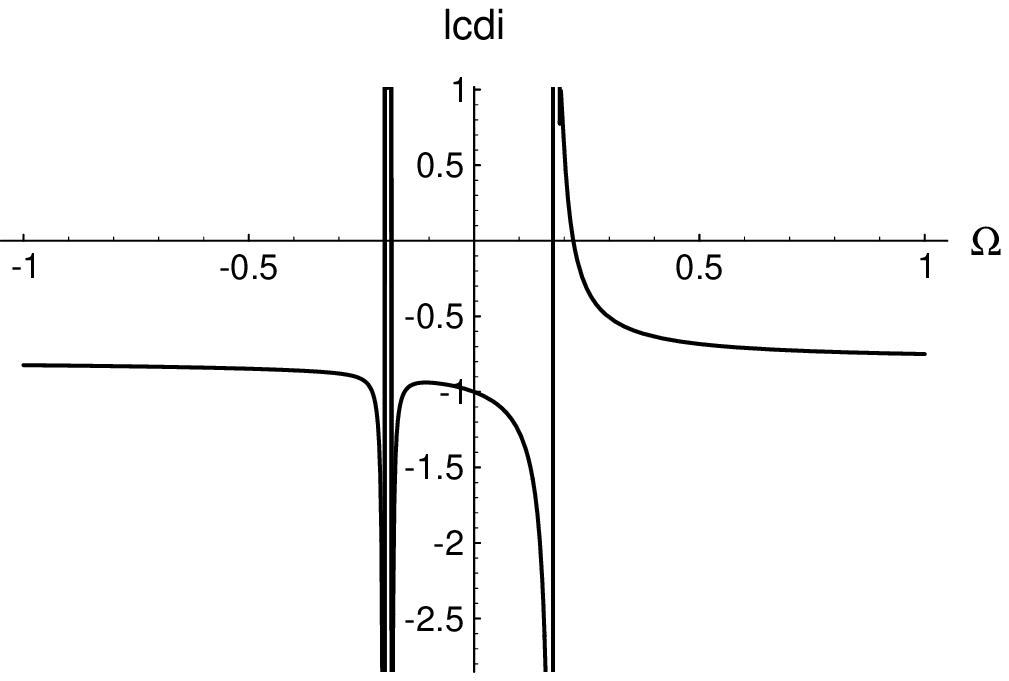}}
\centerline{(b)}
\epsfxsize 3in
\centerline{\epsfbox{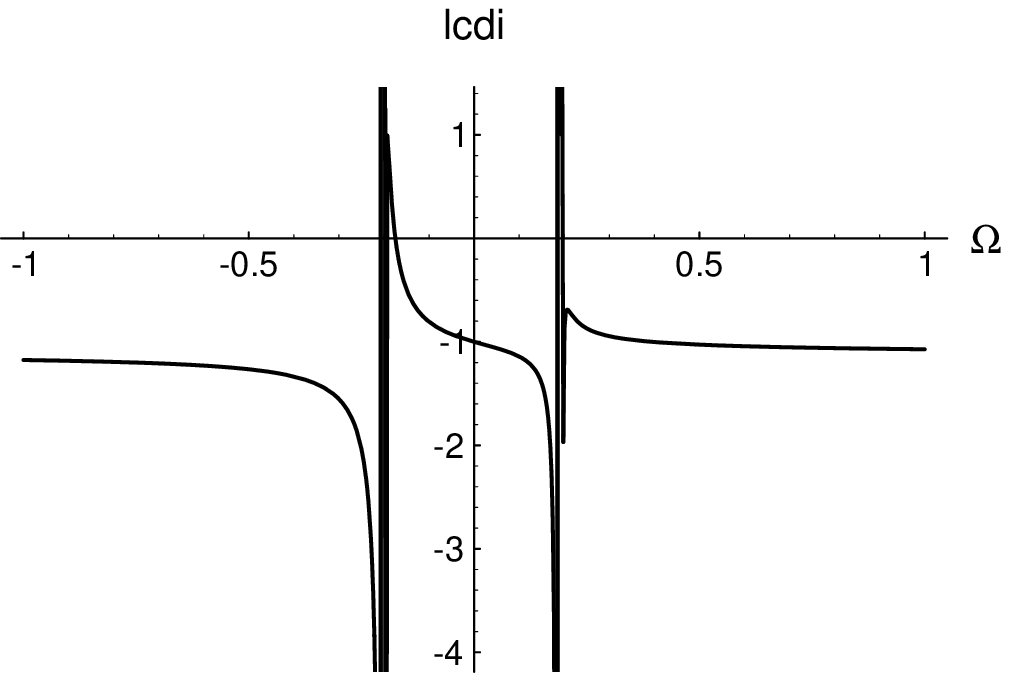}}
\centerline{(c)}
\end{figure}

\begin{figure}
\caption{$N\cal C$ as a function of $\Omega$ (a, b) and as a function of 
$r$ (c, d).} 
\label{fig:ncdi-om}
\begin{tabular}{c@{\hspace{0.5in}}c}
\epsfxsize 2.5in\epsfbox{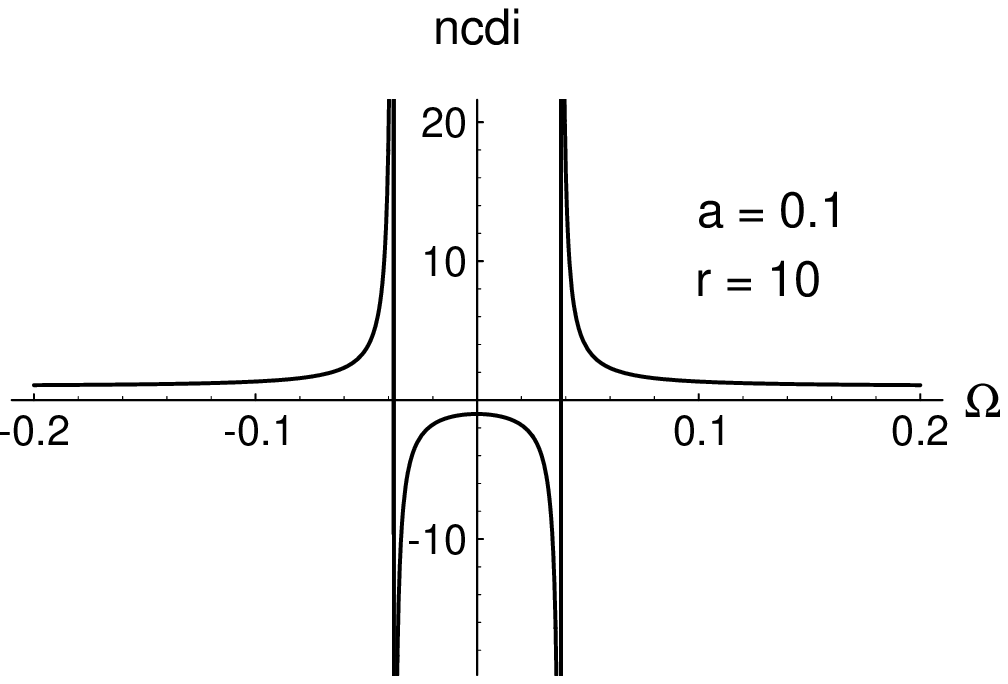}&\epsfxsize 2.5in\epsfbox{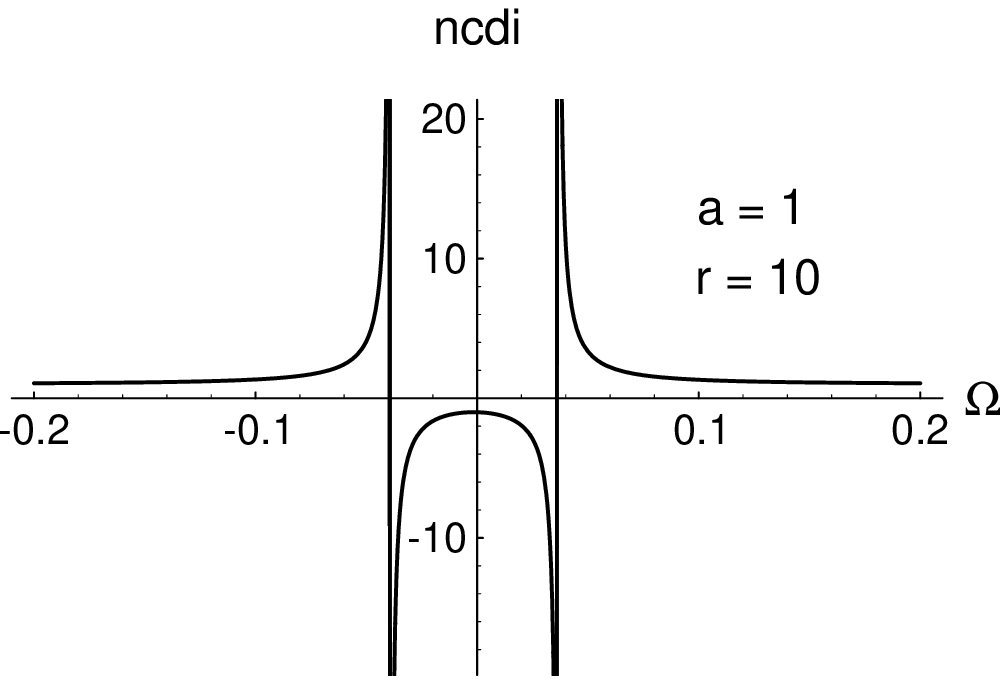}\\
(a)&(b)\\
\epsfxsize 2.5in\epsfbox{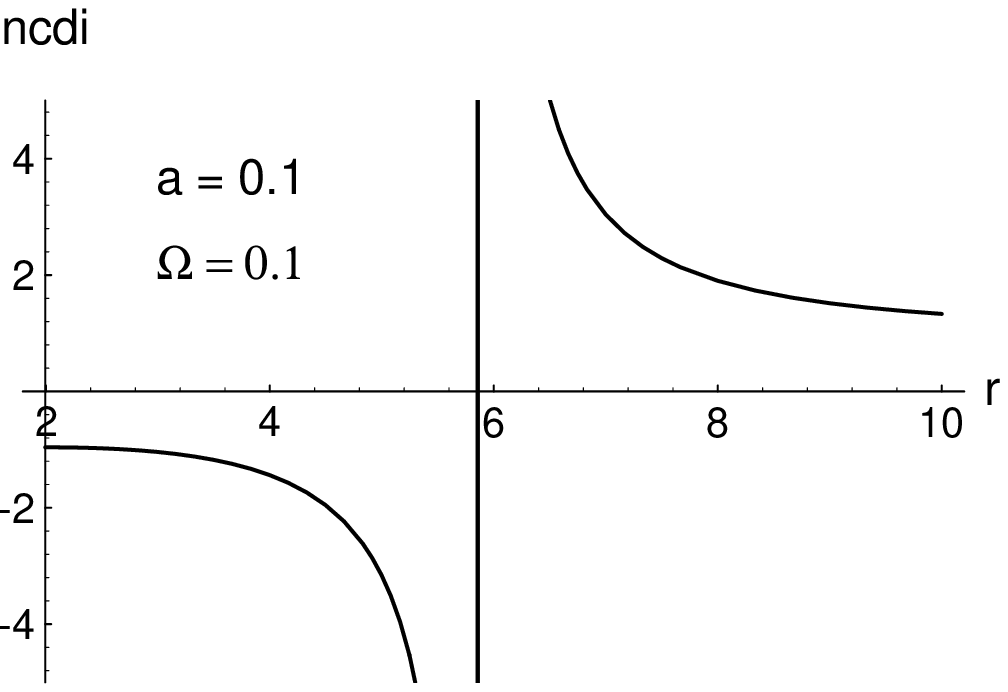}&\epsfxsize 2.5in\epsfbox{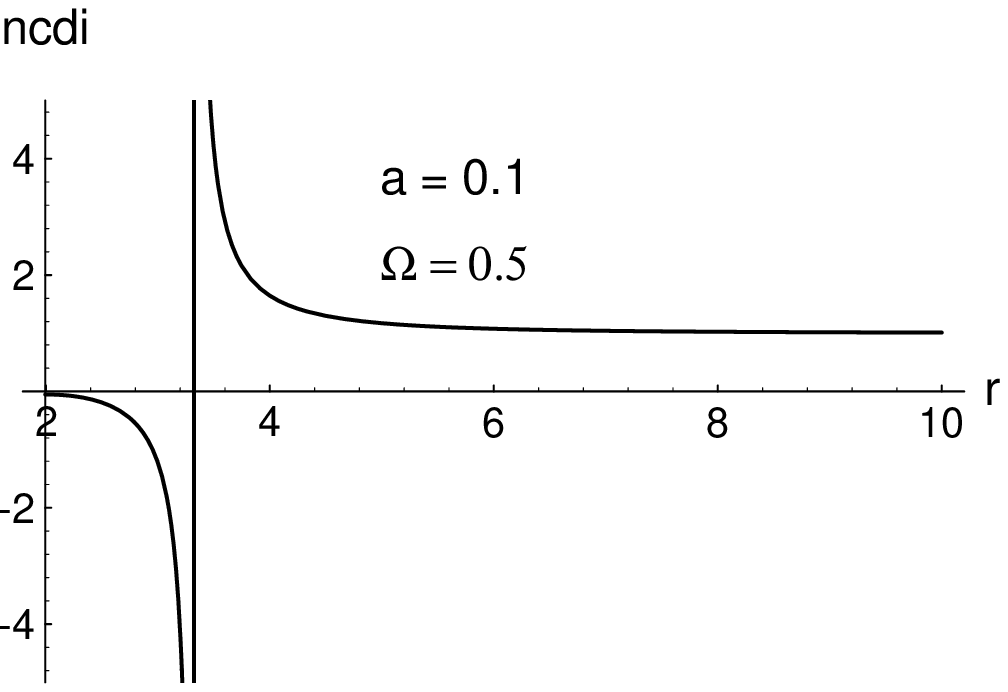}\\
(c)&(d)
\end{tabular}
\end{figure}

\begin{figure}
\caption{$\cal C$ as a function of $\Omega$ for a rotating disk of dust}
\label{fig:dcdi-om}
\begin{tabular}{c@{\hspace{0.5in}}c}
\epsfxsize 2.5in\epsfbox{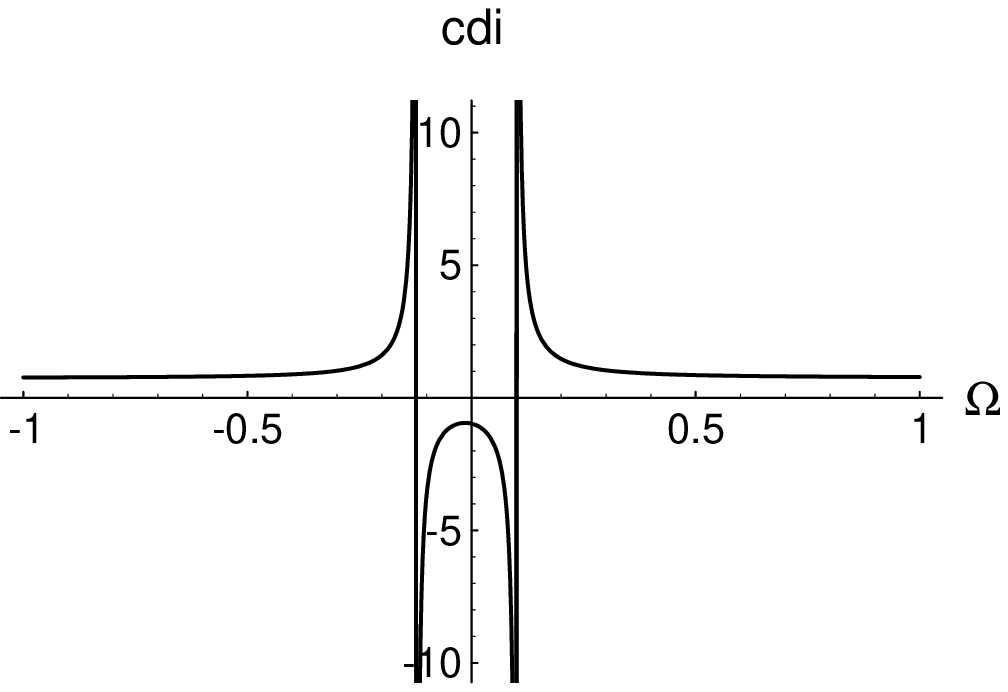}&\epsfxsize 2.5in\epsfbox{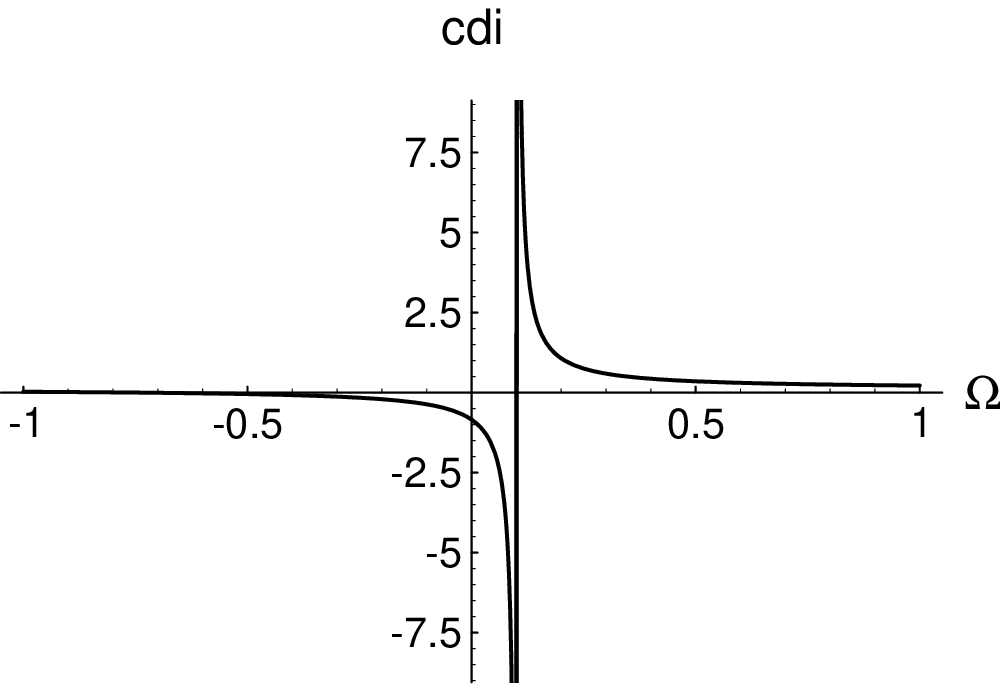}\\
(a)&(b)\\
\epsfxsize 2.5in\epsfbox{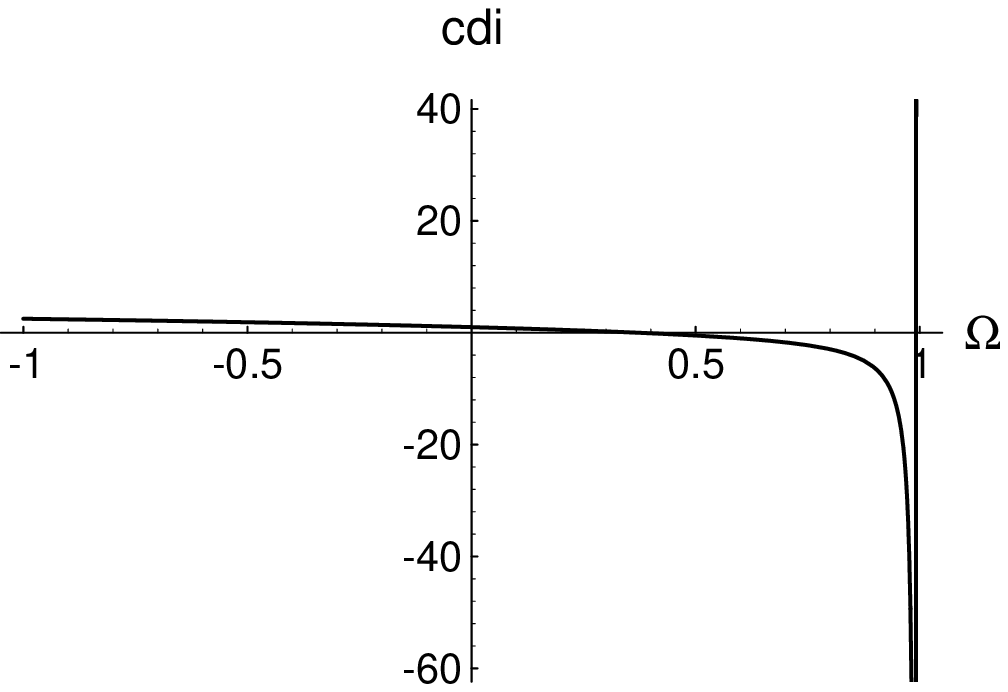}&\epsfxsize 2.5in\epsfbox{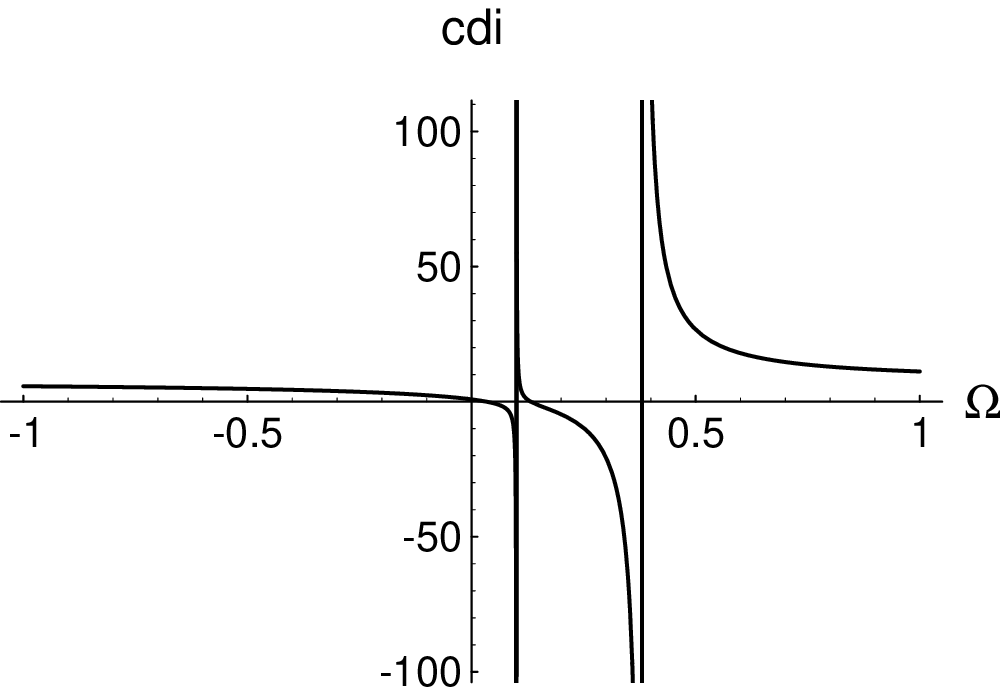}\\
(c)&(d)
\end{tabular}
\end{figure}

\begin{figure}
\caption{Comparison of $\cal C$ and $N\cal C$ for Kerr (a, b) and
the rotating disk (c, d).}
\label{fig:cdi-ncdi}
\begin{tabular}{c@{\hspace{0.5in}}c}
\epsfxsize 2.5in\epsfbox{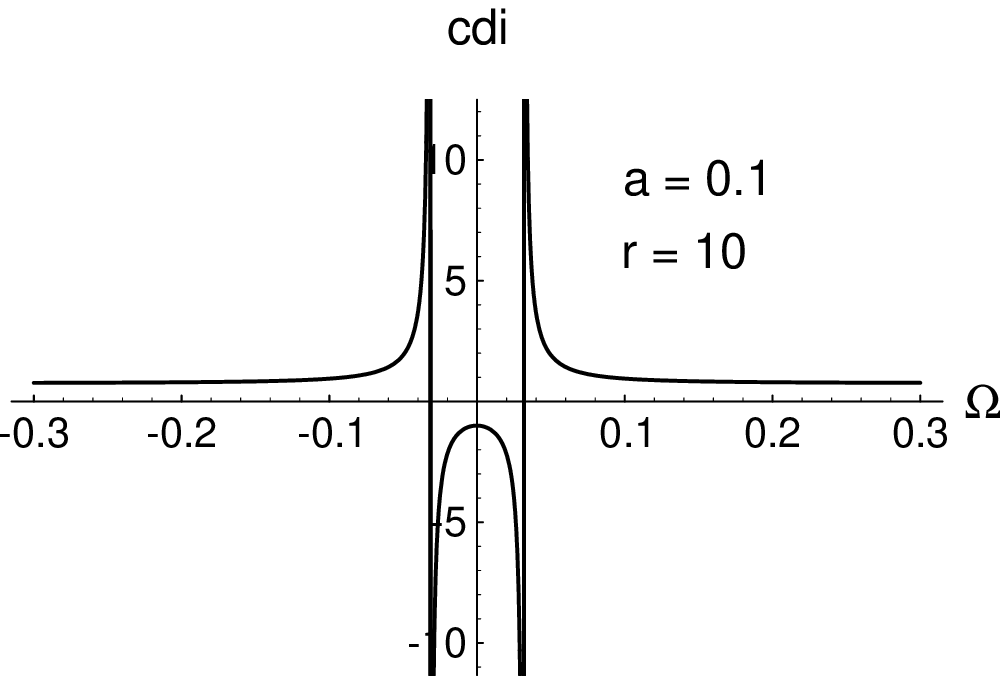}&\epsfxsize 2.5in\epsfbox{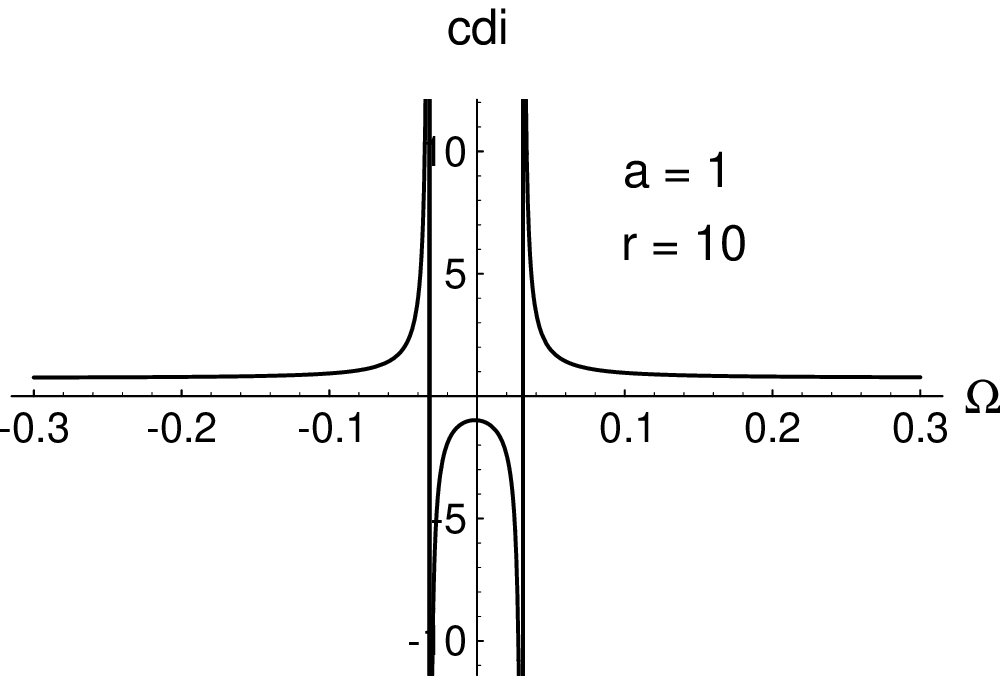}\\
(a)&(b)\\
\epsfxsize 2.5in\epsfbox{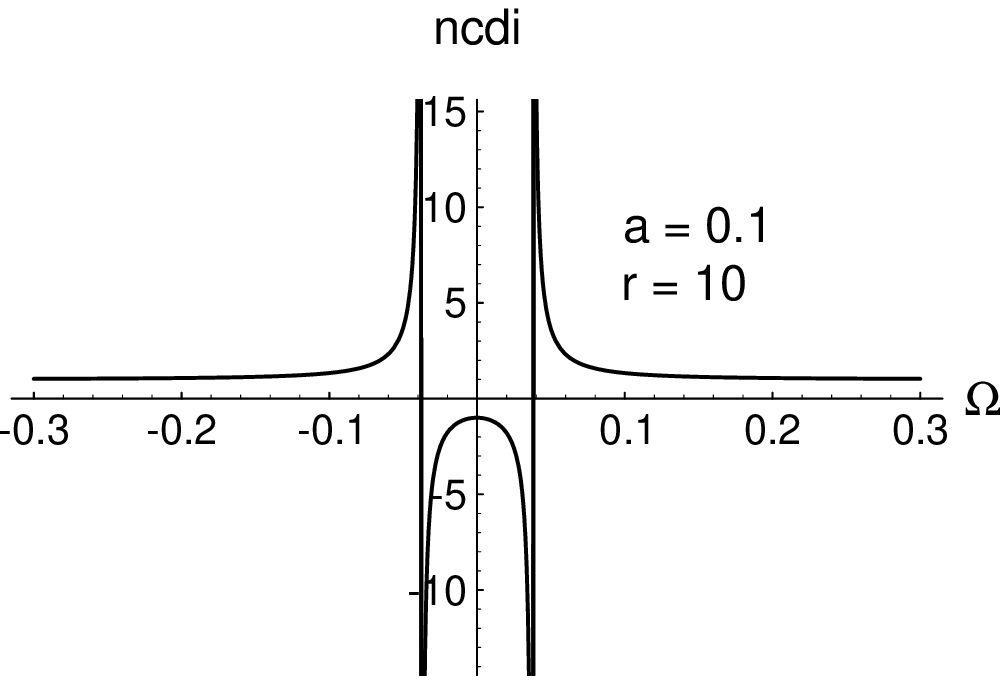}&\epsfxsize 2.5in\epsfbox{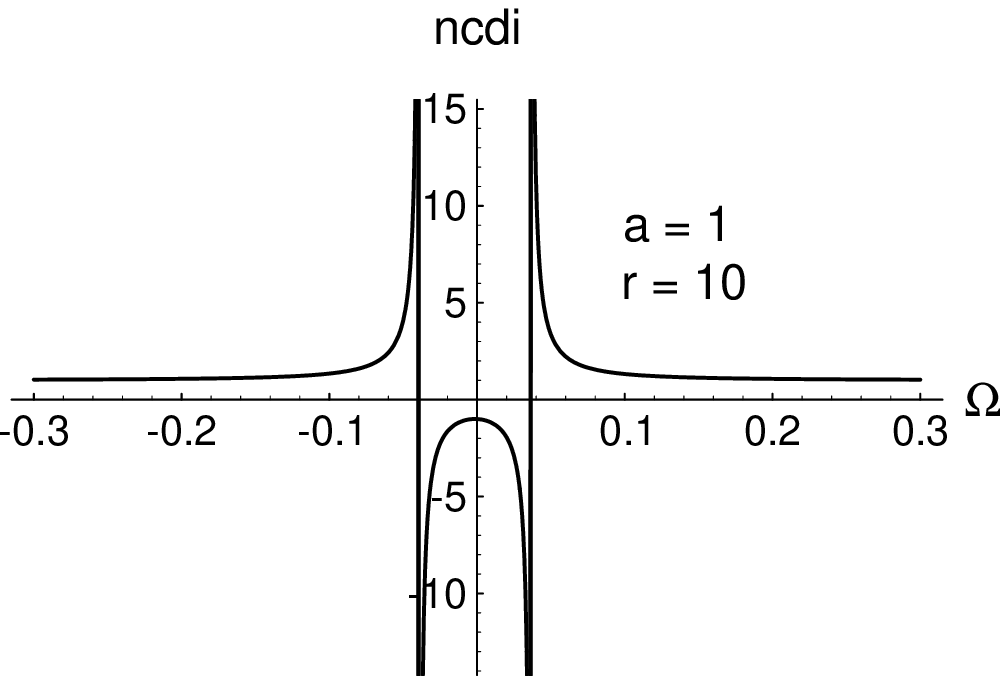}\\
(c)&(d)
\end{tabular}
\end{figure}

\newpage
\centerline{\Large\bf Figure Captions}
\vspace{1in}
\begin{itemize}
\item[\bf Figure \ref{fig:cdi-om}:] $\cal C$ (cdi) as a function of $\Omega$
  ($a=0.5m$) at the prograde photon orbit (a), the retrograde photon
  orbit (b), and the orbit with ${\cal F}_{\rm Cf}=0$ (c).
\item[\bf Figure \ref{fig:lcdi-om}:] $\ell\cal C$ as a function of
  $\Omega$ ($a=0.5m$) at the prograde photon orbit (a), the retrograde
  photon orbit (b), and the orbit with ${\cal F}_{\rm Cf}=0$ (c).
\item[\bf Figure \ref{fig:ncdi-om}:] $N\cal C$ as a function of
  $\Omega$ for fixed $r$ and $a$ (a, b), and as a function of $r$ for
  fixed $a$ and $\Omega$ (c, d).
\item[\bf Figure \ref{fig:dcdi-om}:] $\cal C$ as a function of 
  $\Omega$ for a rotating disk of
  dust with (a) $\mu=0.2$, $\Omega_d=0.1$, (b) $\mu=0.5$,
  $\Omega_d=0.1$, (c) $\mu=2$, $\Omega_d=0.99$ and (d) $\mu=1.9$,
  $\Omega=0.1$.
\item[\bf Figure \ref{fig:cdi-ncdi}:] Comparison of $\cal C$ and
  $N\cal C$ for Kerr (a, b) and the rotating disk (c, d).
\end{itemize}
\newpage
\pagestyle{empty}
\topmargin -0.5in
\newpage
\newpage
\newpage
\newpage
\end{document}